\documentclass[11pt]{article}
\usepackage{amsmath, amsfonts, setspace, url, verbatim, xcolor, cite, graphicx, todonotes, cancel, slashed, mathtools}
\usepackage[T1]{fontenc}
\usepackage[utf8]{inputenc}
\usepackage[
colorlinks=true, 
linkcolor=vub,   
citecolor=vub,   
filecolor=magenta, 
urlcolor=cyan,
]
{hyperref}
\usepackage[a4paper,top=2cm,left=2.5cm,right=2.5cm,bottom=2cm]{geometry}
\usepackage{microtype}
\newcommand{\be}{\begin{equation}}
\newcommand{\ee}{\end{equation}}
\numberwithin{equation}{section}

\newcommand{\hrho}{\hat{\rho}}

\newcommand{\gM}{\mathcal{M}}

\definecolor{vub}{RGB}{0,52,154}
\definecolor{vubo}{RGB}{255,102,0}
\definecolor{redd}{RGB}{255,40,40}
\definecolor{r}{RGB}{228,32,20}
\definecolor{o}{RGB}{238,69,4}
\definecolor{y}{RGB}{253,228,1}
\definecolor{g}{RGB}{108,160,0}
\definecolor{b}{RGB}{0,162,203}
\definecolor{i}{RGB}{120,42,117}

\newcommand{\hr}{\hat{r}}
\newcommand{\htt}{\hat{t}}

\newlength{\bibitemsep}
\newlength{\bibparskip}
\let\oldthebibliography\thebibliography
\renewcommand\thebibliography[1]{%
\oldthebibliography{#1}%
\setlength{\parskip}{\bibitemsep}%
\setlength{\itemsep}{\bibparskip}%
}
\begin{document}

\begin{center}
	\baselineskip=16pt
	
	\vspace{-5cm}
	  
	\hfill IFT-UAM/CSIC-22-126 \\
	
	\vskip 0.5cm

	{\Large \bf  
	Generalised U-dual solutions via $\text{ISO}(7)$ gauged supergravity
	}
	\vskip 2em
	{\large \bf  Chris D. A. Blair{}$^{\,a,b}$, Sofia Zhidkova{}$^b$}
	\vskip 0.6em
	{\it  
	 ${}^a$ Departamento de Física Teórica and Instituto de Física Teórica UAM/CSIC, 
	 Universidad Autónoma de Madrid, Cantoblanco, Madrid 28049, Spain\\
			${}^b$ Theoretische Natuurkunde, Vrije Universiteit Brussel, and the International Solvay Institutes, \\ Pleinlaan 2, B-1050 Brussels, Belgium 
			\\ {\tt c.blair@csic.es}, {\tt Sofia.Zhidkova@vub.be}
	}
	\vskip 0.5cm 
\end{center}

\begin{abstract}
We study a solution generating technique in supergravity which can be viewed as a generalised version of U-duality, taking solutions of type IIA supergravity on a 6-sphere with RR flux to new solutions of 11-dimensional supergravity. The new solutions are characterised by an underlying 6-algebra structure. Our construction provides an 11-dimensional uplift of four-dimensional ISO(7) gauged supergravity, using $E_7$ exceptional geometry techniques. We focus on an example where we start with the D2 solution in type IIA supergravity and construct a new $\tfrac12$-BPS 11-dimensional solution.
\end{abstract}

\tableofcontents

\section{Introduction}

The T- and U-duality symmetries of supergravity act on spacetimes with abelian isometries. A first version of a generalised duality is non-Abelian T-duality (NATD) \cite{delaOssa:1992vci}, which provides a mechanism that dualises a space with non-Abelian isometries to a space with fewer isometries. Both abelian and non-abelian T-duality are special cases of the Poisson-Lie T-duality \cite{Klimcik:1995dy,Klimcik:1995ux}, which can be applied to backgrounds lacking isometries, and which are characterised by an underlying double algebra structure called the Drinfeld double algebra. Further extension of these dualities leads to notions of generalised U-duality, originally proposed using a generalised geometric approach (building on \cite{Hassler:2017yza,Demulder:2018lmj} in the T-duality case) to describe the background, and generalises the Drinfeld double algebra to the so-called exceptional Drinfeld algebra (EDA), that generically is a Leibniz algebra instead of a Lie algebra \cite{Sakatani:2019zrs,Malek:2019xrf,Malek:2020hpo,Sakatani:2020wah}.

In our earlier papers \cite{Blair:2020ndg,Blair:2022gsx} we used this approach to study an attractive example of a generalised U-duality solution generating construction based on the $Sl(5)$ U-duality group acting in four dimensions. 
The relevant exceptional Drinfeld algebra was the Lie algebra $\text{ISO}(4)$.
The generalised U-duality map took solutions of type IIA supergravity on a three-sphere with NSNS flux to new solutions of eleven-dimensional supergravity: a basic example was provided starting with the near horizon NS5 brane.

In this paper, we revisit the generalised U-duality on another example based on the $E_{7}$ U-duality group acting in seven dimensions, with the relevant EDA now being an extension of the $\text{ISO}(7)$ Lie algebra.
We take a near horizon D2 brane solution as a test example, and show how to transform this into a new supergravity solution in eleven dimensions.

The appearance of ISO(4) and ISO(7) algebras is not a choice made a priori but a consequence of choosing to study particular natural algebraic structures, which appear in the definition of the underlying Drinfeld double algebra. 
First of all, we were motivated by the fact that in solutions obtained by NATD, the breaking of translational isometries in the new dual directions can be linked to the appearance of `dual' Lie algebra structure constants $\tilde f^{ab}{}_c$.\footnote{In Poisson-Lie T-duality more generally, these can be interpreted as a coycle of a physical Lie algebra, which in this case is trivial.}
For the dual of $\mathrm{SU}(2)$ i.e. NATD on $\text{S}^3$, these are $\tilde f^{ab}{}_c = \epsilon^{ab}{}_c$. 

In the exceptional Drinfeld algebra \cite{Sakatani:2019zrs,Malek:2019xrf,Malek:2020hpo,Sakatani:2020wah} these dual structure constants are generalised to 3- and 6-algebra structure constants, $\tilde f^{abc}{}_d$ and $\tilde f^{abcdef}{}_g$.\footnote{With interpretations as $n$-cocycles of a physical Lie algebra, which again will be trivial in our examples.}
In the four-dimensional $Sl(5)$ case only the former appear. 
Choosing $\tilde f^{abc}{}_d = \epsilon^{abc}{}_d$ ($a=1,\dots,4$) produced the $\text{ISO}(4)$ algebra studied in \cite{Blair:2020ndg,Blair:2022gsx}. 
The solutions obtained could be seen to directly generalise many of the properties of the solutions resulting from NATD.

In this paper we generalise to the seven-dimensional case, where in principle we can have both the 3- and 6-algebra structures. We choose $\tilde f^{abc}{}_d = 0$ and take $\tilde f^{abcdef}{}_g = \epsilon^{abcdef}{}_g$, which as we explicitly show corresponds to the $\text{ISO}(7)$ algebra.

This ISO(7) example can be viewed as being a sort of electromagnetic dual of our previous ISO(4) case.
This is reflected in the replacement of the 3-algebra structure constants with 6-algebra structure constants, explicitly linked numerologically to the three-form and its magnetic dual six-form, and in the natural choices of NS5 brane (M5 brane on a circle) and D2 brane (M2 brane) as starting points for the construction.

Our approach to constructing new solutions relies on the fact that the generalised geometric realisation of the exceptional Drinfeld algebra provides a mechanism for carrying out a consistent truncation from 10- or 11-dimensional supergravity to a lower-dimensional gauged supergravity. Such truncations allow for both reduction and uplift of solutions. The algebra that is gauged is exactly the EDA. 
When a different consistent truncation is known leading to the same lower-dimensional theory, we can apply `generalised U-duality' by mapping solutions to solutions by reducing via one consistent truncation and uplifting via the other. 
The example of \cite{Blair:2022gsx} gave a consistent truncation of eleven-dimensional supergravity to seven-dimensional ISO(4) gauged maximal supergravity, distinct from the previously known origin of this theory via consistent truncation of type IIA supergravity on a three-sphere with NSNS flux.

In this paper, we will play the same game using reduction and uplift by inequivalent consistent truncations leading to the four-dimensional ISO(7) gauged maximal supergravity.
The first known consistent truncation in this case is provided by type IIA SUGRA on $S^{6}$ \cite{Hull:1988jw,Guarino:2015jca,Guarino:2015vca}.
We apply our solution generating technique by taking any solution of type IIA fitting into the appropriate reduction ansatz, consistently truncating it to a 4-dimensional solution, and then uplifting it to a new 11-dimensional SUGRA solution using the $E_7$ generalised geometry formulation based on the EDA \cite{Sakatani:2020wah}. 
It follows that this method gives an alternative consistent truncation, starting with eleven-dimensional supergravity and leading to $\text{ISO}(7)$ gauged supergravity in four dimensions. 

In fact, this alternative consistent truncation was identified in the paper \cite{Inverso:2017lrz} (which indeed demonstrated the existence of inequivalent consistent truncations for CSO gaugings more generally).
Here we extend, or use, the observation of \cite{Inverso:2017lrz} in the following ways.
Firstly we demonstrate explicitly how to use these inequivalent consistent truncations to perform a generalised U-duality, and explicitly produce a new 11-dimensional supergravity solution using this approach.
We further highlight the algebraic interpretation of the second consistent truncation, by concretely connecting it to the EDA proposal with accompanying $n$-algebra structure, and by comparison to our previous papers \cite{Blair:2020ndg, Blair:2022gsx} we demonstrate how this all fits into the pattern of generalised dualities naturally extending non-abelian T-duality of a three-sphere.

In this paper we specifically apply the uplift procedure to produce a new 11-dimensional solution starting with an extremal D2 brane solution after taking the near horizon limit. Then we analyse the properties of the new 11-dimensional solution, which turn out to be as follows:
\begin{itemize}
\item The new solution can be described by using the generalised geometry techniques with a 6-vector linear in the dual 4-dimensional coordinates. {\itshape (See sections \ref{frame} and \ref{gettingnewsolution}.)}

\item The new solution can be viewed as carrying an electric (M2) charge. {\itshape (See section \ref{charges}.)}

\item The new solution can be viewed as a warped product of $AdS_{4}$, $S^{6}$ and an interval, and it possesses a $\frac{1}{2}$-BPS solution of the 11-dimensional Killing spinor equation.  {\itshape (See section \ref{susy}.)} 

\end{itemize}

In section \ref{algebra} we review the $\text{ISO}(7)$ subalgebra of the $E_{7}$ Drinfeld algebra that we will use in our solution. In section \ref{frame} we construct the frame fields of $E_{7}$ Drinfeld subalgebra.
Then, in section \ref{three} we show an example of how to obtain a new 11-dimensional solution using this technology. In subsection \ref{D2} we start with the initial $D2$ brane solution that we use as an example  of non-vacuum type IIA SUGRA solution. After that, in subsection \ref{gettingnewsolution} we write down the scalar matrix that we take to uplift the initial $D2$ brane solution and construct the new uplifted 11-dimensional SUGRA solution. Then, in sections \ref{charges} and \ref{susy} we describe the properties of the uplifted solution, its charges, local vs global nature, and the amount of supersymmetry it possesses.
We conclude with some brief discussion in section \ref{concl}.

\section{$\text{ISO}(7)$ exceptional Drinfeld algebra and generalised frame}

\subsection{The algebra}\label{algebra} 

The whole $E_{7}$ exceptional Drinfeld algebra was described in \cite{Sakatani:2020wah}.
The 56 generators of the $E_{7}$ exceptional Drinfeld algebra are denoted $T_A =( T_{a}$, $T^{a_{1}a_{2}}$,  $T^{a_{1}...a_{5}}$, $T^{a_{1}...a_{7},a^{\prime}})$, where the Latin indices run from 1 to 7 and sets of multiple indices $a_1 \dots a_p$ are understood to be antisymmetric.
The (generically non-antisymmetric) brackets of these generators can be written generally as:
\be
[T_A, T_B] = X_{AB}{}^C T_C \,.
\label{bracketsX}
\ee
The EDA structure constants $X_{AB}{}^C$ are specified in terms of structure constants $f_{ab}{}^c$, $f^{a_1 \dots a_3}{}_b$, $f^{a_1\dots a_6}{}_b$ and $Z_a$. 
The former three can be formally associated with Lie algebra, 3-algebra and 6-algebra structures. 
In this paper, we focus on non-zero 6-algebra structure constants only, $f^{ a_1 \dots a_6 }{}_b\neq 0$, in which case the algebra is given by the following non-zero brackets:
\begin{equation}
[T_{a},T^{b_{1}...b_{5}}]=-f^{b_{1}...b_{5}c}{}_{a}T_{c}, \hspace{0.4cm} [T_{a},T^{b_{1}...b_{7},b^{\prime}}]=7f^{[b_{1}...b_{6}}{}_{a}T^{b_{7}]b^{\prime}}
\end{equation}
\begin{equation}
[T^{a_{1}...a_{5}},T_{b}]=f^{a_{1}...a_{5}c}{}_{b} T_{c}, \hspace{0.4cm} [T^{a_{1}...a_{5}},T^{b_{1}b_{2}}]=2f^{a_{1}...a_{5}[b_{1}}{}_c T^{b_{2}]c}
\end{equation}
\begin{equation}
[T^{a_{1}...a_{5}},T^{b_{1}...b_{5}}]=-5f^{a_{1}...a_{5}[b_{1}}{}_{c}T^{b_{2}...b_{5}]c}
\end{equation}
\begin{equation}
[T^{a_{1}...a_{5}},T^{b_{1}...b_{7},b^{\prime}}]=-7f^{a_{1}...a_{5}[b_{1}}{}_{c}T^{b_{2}...b_{7}]c,b^{\prime}}-f^{a_{1}...a_{5}b^{\prime}}{}_{c}T^{b_{1}...b_{7},c}
\end{equation}
\begin{equation}
[T^{a_{1}...a_{7},a^{\prime}},T_{b}]=-21f^{[a_{1}...a_{6}}{}_{c}\delta_{bd_{1}d_{2}}^{a_{7}]a^{\prime}c}T^{d_{1}d_{2}}, 
\hspace{0.4cm} [T^{a_{1}...a_{7},a^{\prime}},T^{b_{1}b_{2}}]=7f^{[a_{1}...a_{6}}{}_{c}T^{a_{7}]a^{\prime}cb_{1}b_{2}}
\end{equation}
\begin{equation}
[T^{a_{1}...a_{7},a^{\prime}},T^{b_{1}...b_{5}}]=21f^{[a_{1}...a_{6}}{}_{c}\delta^{a_{7}]a^{\prime}c}_{d_{1}d_{2}e}T^{b_{1}...b_{5}d_{1}d_{2},e}
\end{equation}
In the absence of the other structure constants, the 6-algebra structure constants must obey the identity
\be
f^{da_1\dots a_5}{}_c f^{b_1 \dots b_6}{}_d - 6 f^{a_1 \dots a_5[b_1}{}_d f^{b_2 \dots b_6]d}{}_c = 0 \,,
\label{6jac}
\ee 
ensuring closure of the algebra.
This can be viewed as a generalisation of the Jacobi identity for Lie algebras and the fundamental identity for 3-algebras.

We now further restrict to the following special case:
\be
f^{b_1 \dots b_6}{}_a = \epsilon^{b_1 \dots b_6 c} \delta_{ac} 
\label{specialcase}
\ee
where $\epsilon^{b_{1}...b_{6}c}$ is a 7-dimensional Levi-Civita symbol and $\delta_{ab}$ is seven-dimensional identity matrix.
This is easily verified to obey \eqref{6jac}.
After defining the dualised notations
\begin{equation}
\tilde{T}^{a}=\frac{1}{7!}\epsilon_{a_{1}...a_{7}}T^{a_{1}...a_{7},a}, \hspace{0.3cm} \tilde{T}_{bc}=\frac{1}{5!}\epsilon_{bca_{1}...a_{5}}T^{a_{1}...a_{5}}\,,
\end{equation}
the non-trivial brackets of the algebra then simplify to 
\begin{align} \label{whole}
[T_{a},\tilde{T}_{bc}]=2\delta_{a[b}T_{c]}, & \hspace{0.3cm} [T_{a},\tilde{T}^{b}]=-\delta_{ac}T^{bc}\,, \nonumber \\
[\tilde{T}_{bc},T_{a}]=-2\delta_{a[b}T_{c]}, & \hspace{0.3cm} [\tilde{T}_{ab},T^{cd}]=-4\delta_{e[a}\delta_{b]}^{[c}T^{d]e}\,, \\
[\tilde{T}_{ab},\tilde{T}_{cd}]=4\delta_{\Big[a[c}\tilde{T}_{d]b\Big]}, & \hspace{0.3cm} [\tilde{T}_{ab},\tilde{T}^{c}]=2 \delta_{d[a} \tilde{T}^d \delta_{b]}^{c} \,.\nonumber
\end{align}
The generators $(T_a, \tilde T_{bc})$ generate the $\text{ISO}(7)$ Lie algebra.\footnote{This can be generalised by replacing $\delta_{ab}$ in \eqref{specialcase} by a symmetric matrix of indefinite signature, which would correspond to the algebra of the $\mathrm{CSO}(p,q,r+1)$ gaugings with $p+q+r=7$; replacing $\delta_{ab}$ by a matrix with both symmetric and antisymmetric parts would give something more exotic in which the 28-dimensional `electric' algebra is no longer Lie.}
The other brackets (note that these are not antisymmetric and e.g. $[\tilde T^a, T_b]=0$) match those specified by the $\text{ISO}(7)$ gauging of four-dimensional maximal supergravity (for example, compare with appendix C of \cite{Guarino:2015qaa} where the full structure constants $X_{AB}{}^C$ appearing in \eqref{bracketsX} are given).

\subsection{The generalised frame}\label{frame}

Given any exceptional Drinfeld algebra, a generalised frame can be constructed realising the algebra under the generalised Lie derivative of the appropriate exceptional generalised geometry.
This explicit construction is described in \cite{Sakatani:2019zrs,Malek:2019xrf,Malek:2020hpo,Sakatani:2020wah}.
The data that enters the generalised frame consists of a (left- or right-)invariant vielbein $e^a{}_m$, obeying the Maurer-Cartan equation with Lie algebra structure constants $f_{ab}{}^c$, a 3-vector $\pi^{b_{1}b_{2}b_{3}}$ and a 6-vector $\pi^{b_{1}...b_{6}}$, as well as a scalar function $\Delta$.
The vielbein is linked to a group manifold and the $n$-vectors and scalar obey equations of the form:
\be
\begin{split}
D_{a}\pi^{b_{1}b_{2}b_{3}}&=f^{ b_{1}b_{2}b_{3}}{}_a+\ldots\,,
\\
D_{a}\pi^{b_{1}...b_{6}}&=f^{ b_{1}...b_{6}}{}_a-10f^{[ b_{1}b_{2}b_{3}}{}_a\pi^{b_{4}b_{5}b_{6}]}+\ldots\,,
\\ 
D_a \Delta& = Z_a \,,
\end{split}
\ee
where $D_a \equiv e_a{}^i \partial_i$ and the $\ldots$ corresponds to the terms with Lie algebra structure constants, which are absent in our case.

Now let's construct the necessary data and generalised frame fields for the $E_{7}$  subalgebra with only the six-algebra structure constants $f^{ b_{1}...b_{6}}{}_{a}$ non-trivial. 
The above differential equations then yield $e^{a}_{m}=\delta_{m}^{a}$, $\pi^{b_{1}b_{2}b_{3}}=0$, $\Delta =1$ and allow for a six-vector linear in the coordinates, $\pi^{b_{1}...b_{6}}=x^{i} \delta_i^a f^{ b_{1}...b_{6}}{}_a$.
Then, referring to eq. (5.34) of \cite{Sakatani:2020wah}, we can construct the generalised frame,
which will by definition obey \be
\mathcal{L}_{E_A} E_B = - X_{AB}{}^C E_C 
\ee
under the $E_7$ generalised Lie derivative, thereby realising the algebra of the $\text{ISO}(7)$ gauging.
A generalised frame for the $E_7$ generalised geometry gives a basis $E_A{}^M$ for generalised vectors, which correspond to vectors, two-forms, five-forms and seven-forms tensored with one-forms.
In form notation, the EDA generalised frame describing the ISO(7) algebra has the following elements:
\be
\begin{split}
E_a & = ( e_a, 0,0,0) \,,\\
E^{a_1 a_2} & = (0,e^{a_1} \wedge e^{a_2},0,0) \,,\\
E^{a_1 \dots a_5} & = ( - \pi^{b a_1 \dots a_5} e_b ,0 , e^{a_1} \wedge \dots \wedge e^{a_5} , 0 )\,, \\
E^{a_1 \dots a_7,a'} & = (0, - 7  \pi^{[a_1 \dots a_6} e^{a_7]} \wedge e^{a'} , 0 ,( e^{a_1} \wedge \dots \wedge e^{a_7} ) \otimes e^{a'})\,,
\end{split}
\ee
where in particular the vielbein $e_a$ and one-form $e^a$ have trivial components, $e_a{}^i = \delta_a^i$, $e^a{}_i = \delta^a_i$, and $\pi^{a_{1}...a_{6}}=x_b \epsilon^{a_1 \dots a_6 b}$.

It is useful to record an explicit expression for this frame as a 56 $\times 56$ $E_7$ valued matrix.
The natural decomposition of the generalised vector index is $V^M = ( V^m, V_{m_1m_2}, V_{m_1 \dots m_5}, V_{m_1 \dots m_7, m'})$ but it is convenient to dualise the five-form and mixed symmetry components (as with the algebra generators above) such that the seven-dimensional decomposition used is $V^M = (V^m, V_{m_1m_2},V^{m_1m_2}, V_{m'})$. 
Using this convention for both $M$ and $A$ indices we can write the ISO(7) exceptional Drinfeld algebra generalised frame, or rather its inverse which is more useful for our purposes below, as 
\begin{equation}
E_{M}^{\hspace{0.3cm} A}=\begin{pmatrix}
\delta^{a}_{m} & 0 & 0  & 0 \\
0 & 2\delta_{[a_{1}}^{m_{1}}\delta_{a_{2}]}^{m_{2}}  & 0 & 0\\
2x_{[m_1} \delta_{m_{2}]}^{a} & 0 & 2 \delta_{m_{1}}^{[a_{1}}\delta_{m_{2}}^{a_{2}]} & 0 \\
0 & 2 x_{[a_1} \delta^m_{a_2]} & 0 & \delta_{a}^{m}
\end{pmatrix}
\label{eqframe}\,.
\end{equation} 
This generalised frame (which could also have been constructed using the results of \cite{Inverso:2017lrz})
can be used to construct solutions of 11-dimensional supergravity by uplifting solutions of $\text{ISO}(7)$ gauged supergravity.
Given such a solution, depending on four-dimensional coordinates $y$, and given in terms of the four-dimensional metric $g_{\mu\nu}(y)$, the scalar matrix $M_{AB}(y)$, and one-form $\mathcal{A}_\mu{}^A(y)$, a solution to eleven-dimensional supergravity can be constructed by computing the following quantities:
\be
g_{\mu\nu}(y,x) = g_{\mu\nu}(y) \,,\quad
\gM_{MN} (y,x) = E_M{}^A(x) E_N{}^B(x) M_{AB}(y) \,,\quad
\mathcal{A}_\mu{}^M (y,x) = E_A{}^M (x) \mathcal{A}_\mu{}^A (y) \,,
\label{uplift}
\ee
which correspond to the external metric, generalised metric and external one-form of the $E_7$ exceptional field theory/exceptional generalised geometry description of 11-dimensional supergravity in a $4+7$ split \cite{Coimbra:2011ky,Hohm:2013uia}.
Using the known dictionary between this formulation and the standard variables of 11-dimensional supergravity, the uplifted solution can be extracted.
Conversely the ansatz \eqref{uplift} with the generalised frame \eqref{eqframe} specifies the general form (again on making use of the exceptional geometry dictionary) of a consistent truncation from 11-dimensional supergravity to the ISO(7) gauged supergravity. 
This is a standard application of exceptional geometric techniques (see e.g. \cite{Berman:2012uy,Hohm:2014qga}).

Rather than slavishly work out the full explicit details (which we defer for future work), we will illustrate how this uplift mechanism works on an explicit example, in keeping with our motivation in terms of generalised dualities.
A question which needs to be addressed at this point is how to find examples of solutions which we can feed in to this mechanism.
The ISO(7) gauged supergravity has no known vacua, so we need to consider other sorts of solutions.
A natural candidate is that obtained by the near horizon limit of the D2 brane, which gives a domain wall solution in four dimensions \cite{Boonstra:1998mp}.
We now turn to this solution and its transformation to a new eleven-dimensional solution.

\section{New 11-dimensional solution}
\label{three}

\subsection{The initial D2 brane solution}\label{D2}

In our previous study \cite{Blair:2022gsx} of the $\mathrm{ISO}(4)$ exceptional Drinfeld algebra, we constructed an example of generalised U-duality where we started with the near horizon NS5 solution in type IIA, reduced to seven-dimensional ISO(4) gauged supergravity and uplifted using an $Sl(5)$ exceptional Drinfeld algebra frame to eleven dimensions.
Here we will start with the D2 brane solution in type IIA instead, whose near horizon geometry has the appropriate form for the ISO(7) consistent truncation.
Lifting everything to 11-dimensions, this D2 comes from the M2 while the previously considered NS5 comes from the M5. 
Swapping M5 for M2 reflects the fact that on switching from $\mathrm{ISO}(4)$ to ISO(7) we exchange a trivector for a six-vector, mirroring the exchange of the role of the eleven-dimensional three- and six-forms in the M2 and M5 solutions. In other words, we are applying electromagnetic duality to the entirety of our previous generalised U-duality described in \cite{Blair:2022gsx}.

The D2 brane solution in the string frame is:
\begin{equation}
ds^{2}_{S}=H^{-1/2}[-dt^{2}+dy_{1}^{2}+dy_{2}^{2}]+H^{1/2}[dr^{2}+r^{2}d\Omega^{2}_{(6)}]\,,
\end{equation}
\begin{equation}
e^{-2\phi}=H^{-1/2}\,,\quad C_{ty^{1}y^{2}}=H^{-1}-1\,,
\end{equation}
with $H=1+\frac{1}{r^{5}}$.\footnote{Assuming that by choice of units and rescaling of coordinates we can set all constants to 1.}
The Einstein frame metric is:
\begin{equation}
ds_{E}^{2}=H^{-5/8}[-dt^{2}+dy_{1}^{2}+dy_{2}^{2}]+H^{3/8}[dr^{2}+r^{2}d\Omega_{(6)}^{2}]\,.
\end{equation}
To perform the reduction to a four-dimensional solution, we use the ansatz as in \cite{Guarino:2015jca} for a consistent truncation of type IIA SUGRA on $S^6$ in the Einstein frame.
Assuming all the vector fields $A_{(1)}$ and $B_{(2)}$ appearing in the ansatz are turned off, for the metric and dilaton this ansatz has the form: 
\begin{equation}
ds^{2}_{E}=\Delta^{-1}ds_{4}^{2}+g_{mn}dy^{m}dy^{n}
\,,\quad
e^{-\tfrac{3}{2}\phi}=\Delta\mu_{a}\mu_{b}M^{a8,b8}
\end{equation}
where
\begin{equation}
\mu^{a}\mu^b \delta_{ab} =1 \hspace{0.2cm} \text{in $\mathbb{R}^{7}$}, \hspace{0.3cm} \Delta^{2}=\det g_{mn}/\det \hat{g}_{mn}
\end{equation}
and $\hat{g}_{mn}$ is the round $SO(7)$ symmetric metric on $S^6$. The matrix $M^{a8,b8}$ represents a block of the scalar matrix of the four-dimensional theory.
For the D2 solution, we can use the simplified ansatz
\begin{equation}
M^{a8,b8}\equiv \delta^{ab}M\,.
\end{equation}
Then comparing the dilaton forms we find
\begin{equation}
\Delta M=H^{-3/8}\,,
\end{equation}
and comparing the metric ansatz we deduce
\begin{equation}
g_{mn}=H^{3/8}r^{2}\hat{g}_{mn}, \hspace{0.3cm} \Delta=r^{6}H^{9/8}, \hspace{0.3cm} M=r^{-6}H^{-3/2}\,,
\end{equation}
and the 4-dimensional metric is then
\begin{equation}
ds_{4}^{2}=r^{6}H^{1/2}\Big[-dt^{2}+dy_{1}^{2}+dy_{2}^{2}+Hdr^{2}\Big]\,.
\label{4dextract}
\end{equation}
Since in the D2 brane solution we have a 3-form with all external components, we have to match it with a non-trivial external 3-form of the type IIA gauged SUGRA ansatz on $S^{(6)}$. 
This ansatz is: 
\begin{equation}
C_{(3)}=\mu_{I}\mu_{J}\mathcal{C}^{IJ}, \hspace{0.3cm} \text{where} \hspace{0.2cm} \mathcal{C}^{IJ}=C_{ty_{1}y_{2}}^{\hspace{0.6cm} IJ}dt\wedge dy^{1}\wedge dy^{2}
\end{equation}
thus
\begin{equation}
C_{ty^{1}y^{2}}=\mu_{I}\mu_{J}\mathcal{C}_{ty_{1}y_{2}}^{\hspace{0.6cm} IJ}
\end{equation}
Comparing with the D2 solution, it's not hard to see that
\begin{equation}
\mathcal{C}_{ty_{1}y_{2}}^{\hspace{0.6cm} IJ}=\delta^{IJ}(H^{-1}-1)\,.
\end{equation}
Although this three-form appears in the tensor hierarchy of the gauged supergravity, it does not constitute part of the degrees of freedom of the theory which will be uplifted to eleven dimensions.
In 4 dimensions the field strength of this potential is dual to a scalar (which would therefore require a $-1$ form potential) and in fact this field strength can be related to the scalar potential of the theory \cite{Guarino:2015jca,Guarino:2015vca}. It thus serves as part of the definition of the gauged supergravity and not an independent field within it.

\subsection{Uplifting the scalar matrix and obtaining the new solution} \label{gettingnewsolution}

Let's construct the full $56\times 56$ scalar matrix $M_{AB}$ (the flat index $A=(ab,a8)$, where $a$ runs from 1 to 7):
\begin{equation}
M_{AB}=\begin{pmatrix}
 M_{a8, b8}& M_{a8}^{\hspace{0.4cm} cd} & M_{a8, cd}  & M_{a8}^{\hspace{0.4cm} c8} \\
 M^{ab}_{\hspace{0.4cm} cd}& M^{ab, cd} & M^{ab}_{\hspace{0.4cm} c8}  & M^{ab, c8}\\
M_{ab,c8} &M_{ab}^{\hspace{0.4cm} cd} & M_{ab,cd} &   M_{ab}^{\hspace{0.4cm} c8}\\
M^{a8}_{\hspace{0.4cm} cd} & M^{a8, cd} & M^{a8}_{\hspace{0.4cm} b8}  & M^{a8, b8}
\end{pmatrix}\,,
\end{equation}
from which the generalised metric of the eleven-dimensional uplift is constructed as follows
\begin{equation}
\mathcal{M}_{MN}=E_{M}^{\hspace{0.3cm} A}M_{AB}E^{B}_{\hspace{0.3cm} N}\,.
\end{equation}
In order to construct the $M_{AB}$ matrix we refer to the dictionary described in \cite{Guarino:2015jca}, from where, comparing with the form of the $D2$ brane solution of the previous section
\begin{equation}
M_{AB}=\begin{pmatrix}
r^{-4}H^{-1/2}\delta_{ab}  & 0 & 0& 0\\
0 & r^{-8}H^{-3/2}\delta^{a_{1}[b_{1}}\delta^{b_{2}]a_{2}} & 0 & 0 \\
0 & 0 & r^{-2}H^{-1/2}\delta_{a_{3}[b_{3}}\delta_{b_{4}]a_{4}} & 0\\
0 & 0 & 0 & r^{-6}H^{-3/2}\delta^{a_{5}b_{5}}
\end{pmatrix}
\end{equation}
Here to meet the requirement of det$M$$=1$ we have to impose the near-horizon limit of the D2 brane solution by setting $H=\frac{1}{r^{5}}$.

The generalised metric describing the new uplifted solution is, after using the generalised frame \eqref{eqframe}
\begin{equation}
\mathcal{M}_{MN}=\begin{pmatrix}
r^{-3/2} \delta_{mn} & 0 & 2 r^{-3/2}  \delta_{m[n_{2}}x_{n_{1}]} & 0 \\
0 & 2 r^{-1/2} \delta^{m_{1}[n_{1}}\delta^{n_{2}]m_{2}} & 0 & 2 r^{-1/2} x^{[m_{1}}\delta^{m_{2}]n} \\
2 r^{-3/2} \delta_{n[m_{2}}x_{m_{1}]} & 0 & r^{1/2} K_{m_1m_2,n_1n_2}& 0 \\
0 & 2r^{-1/2}\delta^{m [n_2}x^{n_{1}]} & 0 & r^{3/2}K^{m n}
\end{pmatrix}
\label{upliftedgenmet}
\end{equation}
where
\begin{equation}
K_{m_1m_2,n_1n_2}=2\delta_{m_{1}[n_{1}}\delta_{n_{2}]m_{2}}+4r^{-2}x_{[m_{2}}\delta_{m_{1}][n_{1}}x_{n_{2}]}, \hspace{0.3cm} K^{mn}=\delta^{mn}(1+r^{-2}x_{a}x^{a})-r^{-2}x^{m}x^{n}\,,
\end{equation}
We need to compare this with the expression for the parametrisation of the $E_{7}$ generalised metric  in terms of the internal seven-dimensional components of the metric $\phi_{mn}$, three-form and six-form.
Referring for example to \cite{Berman:2011jh}, we see that \eqref{upliftedgenmet} corresponds to a generalised metric with vanishing three-form but non-trivial six-form.
The precise parametrisation of the generalised metric that we need (taking care to follow the conventions of \cite{Sakatani:2020wah} which we used to construct the EDA generalised frame) then has the form:
\begin{equation} \label{ans}
\mathcal{M}_{MN}=\begin{pmatrix}
\phi^{\tfrac12} L_{mn} & 0 & 2\phi_{m[n_{2}}U_{n_{1}]} & 0 \\
0 & \phi^{\tfrac12} ( 2\phi^{m_{1}[n_{1}}\phi^{n_{2}]m_{2}}+4U^{[m_{1}}\phi^{m_{2}][n_{1}}U^{n_{2}]}) & 0 &2\phi^{n[m_{2}}U^{m_{1}]}\\
2\phi_{n[m_{2}}U_{m_{1}]} & 0 & 2\phi^{-\tfrac12}\phi_{m_{1}[n_{1}}\phi_{n_{2}]m_{2}} & 0 \\
0 &2\phi^{m[n_{2}}U^{n_{1}]} & 0 & \phi^{-\tfrac12}\phi^{mn}
\end{pmatrix}
\end{equation}
where $\phi=\det(\phi_{mn})$, 
\begin{equation}
U^{m}=\frac{1}{6!}\phi^{-1/2}\epsilon^{mn_{1}...n_{6}}C_{n_{1}...n_{6}}\,,\quad
L_{mn}\equiv \phi_{mn}(1+U_{p}U^{p})-U_{m}U_{n}\,,
\end{equation}
and $U_m =\phi_{mn} U^n$, where here $\epsilon$ denotes the alternating symbol.

Comparing the two expressions we find that the seven-dimensional internal metric is:
\begin{equation}
\phi_{mn}=r^{-1/3}(1+r^{-2}x_{p}x^{p})^{-1/3}\big[\delta_{mn}+r^{-2}x_{m}x_{n}\big]
\end{equation}
and that the six-form is:
\begin{equation}
C_{m_{1}...m_{6}}=\epsilon_{m_{1}...m_{6}n} x^{n}r^{-2}(1+r^{-2}x_{p}x^{p})^{-1}\,.
\label{sixform}
\end{equation}
The latter gives rise to the field strength components
\begin{equation}\label{fint}
F_{m_{1}...m_{7}}=\epsilon_{m_{1}...m_{7}}r^{-2}(1+r^{-2}x_{n}x^{n})^{-2}\big[7+5 r^{-2}x_{p}x^{p}\big]\,,
\end{equation}
\begin{equation}\label{fmix}
F_{r m_{1}...m_{6}}=-2\epsilon_{m_{1}...m_{6}n}x^{n}r^{-3}(1+r^{-2}x_{p}x^{p})^{-2}\,.
\end{equation}
Now using the ExFT construction we can build the full new 11-dimensional solution.
The 11-dimensional metric is:
\begin{equation}
\hat{g}_{\hat{\mu}\hat{\nu}}=\begin{pmatrix}
|\phi|^{\omega}g_{\mu\nu}^{ExFT} + A_\mu{}^k A_\nu{}^l \phi_{kl}  & A_{\mu}^{k}\phi_{kn} \\
A_{\nu}^{k}\phi_{km} & \phi_{mn}
\end{pmatrix} \,,
\end{equation}
where $\omega=-\frac{1}{n-2}=-\frac{1}{2}$ in our case of $n=11-d=4$.
The 4-dimensional ExFT metric is that extracted in \eqref{4dextract} from the D2 brane solution, in the near horizon limit:
\begin{equation}
(ds^{2})^{ExFT}=r^{7/2}[-dt^{2}+dy_{1}^{2}+dy_{2}^{2}+r^{-5}dr^{2}] \,,
\end{equation}
and as there is no one-form present we have $A_{\mu}^{k}=0$.
Thus, the new 11-dimensional metric is
\begin{equation}
\hat{ds}^{2}_{11}=r^{-1/3}(1+r^{-2}x_{k}x^{k})^{-1/3}\Big[r^{5}(1+r^{-2}x_{p}x^{p})[-dt^{2}+dy_{1}^{2}+dy_{2}^{2}+r^{-5}dr^{2}]+(\delta_{mn}+r^{-2}x_{m}x_{n})dx^{m}dx^{n}\Big]
\end{equation}
The only gauge field components present are those of the six-form given in \eqref{sixform}.
We can rewrite our solution in different coordinate systems.
We can pass to spherical coordinates in place of the $x^i$, 
in terms of which we we can rewrite the new 11-dimensional metric as
\begin{equation}\label{sphrcor}
\hat{ds}^{2}_{11}=r^{1/3}(r^2+\rho^2)^{2/3} \Big[r^{3}\big(-dt^{2}+dy_{1}^{2}+dy_{2}^{2}+r^{-5}dr^{2}\big)+ r^{-2} d\rho^{2} \Big]
+ r^{1/3} (r^2+\rho^2)^{-1/3} \rho^{2}d\Omega_{(6)}^{2}
\end{equation}
where $\rho^2 \equiv x_i x^i$ and $d\Omega_{(6)}^{2}$ denotes the metric on the unit six-sphere.
The six-form potential and its field strength are:
\be
C_{(6)} = \frac{\rho^7}{r^2+\rho^2} \mathrm{Vol}_{S^6} \,,\quad
F_{(7)} = - \frac{2 \rho^7}{(r^2 + \rho^2)^2} r d r \wedge  \mathrm{Vol}_{S^6}  + \frac{7r^2 + 5 \rho^2}{(r^2+\rho^2)^2} \rho^6 d \rho \wedge  \mathrm{Vol}_{S^6} \,.
\ee
The four-form field strength obtained by Hodge dualisation is
\be
F_{(4)} = r^4 ( 7 r^2 + 5 \rho^2) dt \wedge d y^1 \wedge d y^2 \wedge dr + 2 r^5 \rho dt \wedge d y^1 \wedge d y^2 \wedge d\rho \,.
\ee
A further coordinate change relates the 4-dimensional part of the metric to a familiar form of the metric on $AdS_{4}$.
This is a property inherited from the original D2 solution, whose near horizon string frame metric is a function of the radial coordinate times $AdS{}_4 \times S^6$ (in a dual frame \cite{Boonstra:1998mp} the metric is exactly $AdS{}_4 \times S^6$).
By introducing a new coordinate 
\begin{equation}
\tilde{r}\equiv \frac{2}{3}r^{3/2}
\end{equation}
then the 4-dimensional bit of the solution can be shown to involve an $AdS_{4}$ metric in the Poincare patch, using the fact that
\begin{equation}
r^{3}[-dt^{2}+dy_{1}^{2}+dy_{2}^{2}+r^{-5}dr^{2}]=\mathcal{R}^{-2} \tilde{r}^{2}[-dt^{2}+dy_{1}^{2}+dy_{2}^{2}]+\mathcal{R}^2 \frac{d\tilde{r}^{2}}{\tilde{r}^{2}}
\end{equation}
where $\mathcal{R} = 2/3$ is the AdS radius.

We can finally comment on the behaviour of our metric as $r \rightarrow 0$. 
The Ricci scalar is
\be
R = - \tfrac16 r^{-1/3} (49r^2 + 25\rho^2) (r^2 + \rho^2)^{-5/3} 
\ee
and so the solution is singular for $r \rightarrow 0$.
This is also a feature of the D2 brane near horizon solution.

\subsection{Properties of the new solution}

\subsubsection{Charges and global properties} \label{charges}

The solution that we have obtained is a local solution: we have not yet specified the range of the coordinates $x^i$, or alternatively that of $\rho$ if we change to spherical coordinates.
The situation is entirely analogous to that found when obtaining solutions via non-abelian T-duality, and to our previous generalised U-duality construction \cite{Blair:2022gsx}.
If the $x^i$ are to be regarded as periodically identified then our solution can be regarded as a non-geometric background, globally identified up to a non-trivial $E_7$ transformation acting as a constant shift of the six-vector used in constructing the solution, as noted in \cite{Inverso:2017lrz} and similar to examples in \cite{Fernandez-Melgarejo:2017oyu, Bugden:2019vlj,Blair:2022gsx,Blair:2022ahh}.
Alternatively, we can work in the spherical coordinates and attempt to fix the range of $\rho$ by requiring the solution carry well-defined brane charges.

Accordingly, let's consider the charges of the new uplifted solution. 
It only carries electric M2 charge, namely
\begin{equation}
Q_{M2}\sim \int \star F_4 = \int d C_6 \,,
\end{equation}
where from above $C_6 = \rho^7/(r^2+\rho^2) \mathrm{Vol}_{S^6}$.
We could try to specify a seven-cycle to evaluate this charge (generalising the argument of \cite{Terrisse:2018hhf} for non-abelian T-dual solutions) by integrating from $\rho=\rho_0$ to some value $\rho=\rho_1$ at a fixed value of $r= r_0$, and then integrate from $r= r_0$ to $r=r_1$ at fixed $\rho=\rho_1$, such that the six-sphere part of the solution vanishes at $\rho=\rho_0$ and $r=r_1$. 
The result is independent of $r_0$, and gives $16 \pi^3 \rho_1^7/15(r_1^2+\rho_1^2)$.
Choosing $\rho_0=0$ and $r_1=0$ would give an electric charge $Q_{M2} \sim 16\pi^3 \rho_1^5/15$, which on properly reinserting dimensionful constants could be argued to fix $\rho_1$ by requiring the charge is an integer times the M2 charge.

Note that this M2 charge is analogous to the M5 charge appearing in our earlier solution \cite{Blair:2022gsx}, hence in this ``dual'' example the electric and magnetic charges are swapped, mirroring the swap of trivector and six-vector we noted earlier.
To be more specific, the relevant M5 charge of \cite{Blair:2022gsx} is that which is present when the initial solution there is solely the near horizon NS5 brane.
It was also possible in \cite{Blair:2022gsx} to start with an F1-NS5 intersection. The resulting new 11-dimensional solution then required a different global completion which was possible at least for its AdS${}_3$ limit. This limit fit into a class of solutions \cite{Lozano:2020bxo} in a manner reminiscent of AdS solutions obtained via non-abelian T-duality. This involved a linear function of $\rho^2$, defined on a series of subintervals with jumps in slope across each subinterval. It is unclear if it is possible to apply similar thinking to our example in this paper (which has a more complicated functional dependence on the $r$ coordinate alongside $\rho$), or to find or classify other solutions built using the ISO(7) generalised frame.

\subsubsection{SUSY analysis} \label{susy}

Let us now look at the solution of the Killing spinor equation and find out how many supersymmetries the new uplifted solution has.
The Killing spinor equation we need to solve is\footnote{We follow the conventions we used in \cite{Blair:2022gsx}, in particular $\{ \Gamma_a , \Gamma_b \} = 2 \eta_{ab}$ with $\eta_{ab}$ having mostly minus signature.}
\begin{equation}
\delta_\epsilon \psi_{\mu}=2\partial_{\mu}\epsilon-\frac{1}{2}\omega_{\mu}{}^{ ab}\Gamma_{ab}\epsilon+\frac{i}{144}(\Gamma^{\alpha\beta\gamma\delta}_{\hspace{0.8cm} \mu}-8\Gamma^{\beta\gamma\delta}\eta^{\alpha}_{\mu})\epsilon F_{\alpha\beta\gamma\delta}=0
\end{equation}
where the Greek indices are the curved coordinates, and Latin indices are the flat ones. 
For the $t$-component (and similarly for $y^{1}$ and $y^{2}$), using the hatted indices for the curved coordinates, and unhatted for the flat ones, we explicitly have:
\begin{equation}\label{2tt}
\Gamma^{t}\partial_{\htt}\epsilon+\frac{1}{6}r^{1/2}(1+r^{-2}\rho^{2})^{-1}\Big[2\rho\Gamma_{\rho}+7r(1+\frac{5}{7}r^{-2}\rho^{2})\Gamma_{r}-i\Big(2\rho\Gamma^{ty^{1}y^{2}\rho}+7r(1+\frac{5}{7}r^{-2}\rho^{2})\Gamma^{ty^{1}y^{2}r}\Big)\Big]\epsilon=0\,.
\end{equation}
Assuming that $\epsilon$ is $t$-independent (similarly $y^{1}$ and $y^{2}$ independent), and looking at the similar coordinate dependence in front of the same gamma-matrix combination, we can extract the following projection condition on $\epsilon$
\begin{equation}\label{condit1}
(1+i\Gamma^{ty^{1}y^{2}})\epsilon=0
\end{equation}
which we can use in solving the rest of the equations. 

The $r$ and $\rho$ equations become
\begin{equation}
\partial_{\hr}\epsilon=r^{-1}(1+r^{-2}\rho^{2})^{-1}\Big[1+\frac{1}{6}(1+5r^{-2}\rho^{2})-\frac{1}{2}r^{-1}\rho\Gamma_{r\rho}\Big]\epsilon
\end{equation}
\begin{equation}
\partial_{\hrho}\epsilon=\frac{1}{6}r^{-1}(1+r^{-2}\rho^{2})^{-1}[2r^{-1}\rho+3\Gamma_{r\rho}]\epsilon
\end{equation}
with the common solution
\begin{equation}
\epsilon = \epsilon_{r \rho} \bar \epsilon \,,\quad
\epsilon_{r\rho}=r^{7/6}(1+r^{-2}\rho^{2})^{1/6}\exp\left[-\frac{1}{2}\Gamma_{r\rho}\tan^{-1}(\frac{r}{\rho})\right]\,,
\end{equation}
where $\bar \epsilon$ depends on the $S^6$ coordinates only. 
Now, working in round spherical coordinates $(\chi,\theta_1,\dots,\theta_5)$ on $S^6$, we can find a solution of the form $\bar \epsilon = \epsilon_\chi \epsilon_{\theta_1} \dots \epsilon_{\theta_5} \epsilon_0$ with $\epsilon_0$ a constant spinor.
Indeed, we firstly have the equation
\begin{equation}
\partial_{\hat{\chi}}\epsilon=\frac{1}{2}(1+r^{-2}\rho^{2})^{-1/2}[\Gamma_{\rho\chi}+r^{-1}\rho\Gamma_{r\chi}]\epsilon
\end{equation}
where we can commute the gamma matrices from the $\epsilon_{r\rho}$ part, moving it to the left of both sides of the equation, and end up solving for $\epsilon_{\chi}$
\begin{equation}
\epsilon_{\chi}=\exp\big[\frac{1}{2}\Gamma_{r\chi}\chi\big]
\end{equation}
and in a similar manner for the rest of the 5 angles $\theta_{1}....\theta_{5}$ we find
\begin{equation}
\epsilon_{\theta_{1}}=\exp\big[\frac{1}{2}\Gamma_{\chi\theta_{1}}\theta_{1}\big], \hspace{0.3cm} \epsilon_{\theta_{2}}=\exp\big[\frac{1}{2}\Gamma_{\theta_{1}\theta_{2}}\theta_{2}\big], \hspace{0.2cm} \text{etc...}
\end{equation}
so the final solution is of the form
\begin{equation}
\epsilon=\epsilon_{r\rho}\epsilon_{\chi}\epsilon_{\theta_{1}}\cdots\epsilon_{\theta_{5}}\epsilon_{0}
\end{equation}
where after applying the condition \eqref{condit1} $\epsilon_{0}$ is a constant spinor satisfying
\begin{equation}
(1+i\Gamma^{ty^{1}y^{2}})\epsilon_{0}=0
\end{equation}
which kills a half of the total degrees of freedom, thus, our solution is $\frac{1}{2}$-BPS.
This is consistent with the supersymmetry of the initial D2 solution and with the supersymmetry preservation of our previous example of `generalised U-duality' \cite{Blair:2022gsx}.

\section{Conclusion}
\label{concl}
In this paper we discussed another example of a solution generating mechanism which can be viewed as a generalised U-duality transformation.
We used a special case of the $E_{7}$ exceptional Drinfeld algebra, describing the four-dimensional $\text{ISO}(7)$ gauging, and used this to construct a new 11-dimensional solution starting with the near horizon limit of the $D2$ brane solution of type IIA SUGRA. This can be seen as a ``dual'' construction (in the electromagnetic sense) of our previous example, based on the $Sl(5)$ exceptional Drinfeld algebra corresponding to the seven-dimensional $\text{ISO}(4)$ gauging \cite{Blair:2022gsx}.
Together these examples generalise, in a particular manner, features of non-abelian T-duality to the 11-dimensional setting (see table \ref{table}), using the natural exceptional Drinfeld algebra cases with either non-trivial $3$- and $6$-algebra structure constants, and hence non-trivial tri- and six-vectors.

\begin{table}[h]
\centering
\begin{tabular}{ccc}
\underline{Solution obtained by:} & \underline{Algebraic structure} & \underline{Generalised frame} \\ 
Non-abelian T-duality of $S^3$ &  $\tilde f^{ab}{}_c = \epsilon^{ab}{}_c$ & bivector \\
$Sl(5)$ generalised U-duality of $S^3$ (w/NSNS flux) &  $\tilde f^{abc}{}_d = \epsilon^{abc}{}_{d}$ & trivector \\ 
$E_7$ generalised U-duality of $S^6$ (w/RR flux) &  $\tilde f^{abcdef}{}_g =\epsilon^{abcdef}{}_g$  & six-vector
\end{tabular}
\caption{Properties of generalised dualities}
\label{table}
\end{table}

We have so far only considered the M-theory realisation of the EDA, but it would be interesting to systematically explore similar features in its IIA and IIB decompositions. Here we would expect to construct a variety of other generalised frames involving $n$-vectors with a linear coordinate dependence, and identify the lower-dimensional gaugings these capture. 
 
The usefulness of these constructions depends on whether the choice of EDA allows one to access gauged supergravities with either interesting known solutions or known alternative origins as consistent truncations from 10- and 11-dimensions.
In this paper and in \cite{Blair:2022gsx} we used the latter approach to identify brane solutions at the 10-dimensional level to which we could apply reduction and uplift.
The ISO(7) example of this paper is a case where there are in fact no known vacua (the D2 brane solution reducing to a domain wall solution). 
We have made choices for the EDA which seemed algebraically `natural' and to some extent gotten lucky in finding that these corresponded to uplifts of known gauged supergravities in fact corresponding to consistent truncations already identified from a different, though related, perspective in \cite{Inverso:2017lrz}.
It would be good to extend and improve this search strategy, including to situations with simultaneously non-trivial 3- and 6-algebra structure constants, and more broadly to try to understand exactly what is the common feature (spheres with flux?) of the initial solutions `dual' to the solutions built using these EDA generalised frames, and how the $n$-algebra symmetry manifests in these background (if at all).

A natural question about the ISO(7) case concerns whether we can do anything with the \emph{dyonic} ISO(7) gaugings \cite{DallAgata:2011aa,DallAgata:2012mfj,DallAgata:2014tph}, which have a richer vacuum structure.
These gaugings can be obtained by a consistent truncation from the massive type IIA theory on $S^6$ \cite{Guarino:2015jca,Guarino:2015vca}: we have been using this consistent truncation in the massless limit for the ISO(7) gauging. It can be quickly checked that this gauging modifies the algebra \eqref{whole} with additional non-zero brackets including $[\tilde T^a, \tilde T^b] \sim T^{ab}$. This bracket is however always zero in the EDA construction \cite{Sakatani:2020wah}.
Hence the dyonic ISO(7) algebra is not an EDA  -- if it were we would immediately know how to construct a [geometric] generalised frame realising it.
Indeed we have been informed by Y. Sakatani that making this bracket non-zero in an extension of the EDA always requires locally non-geometric R-fluxes, in agreement with the statement of \cite{Inverso:2017lrz} implying the dyonic ISO(7) gauging does not admit a locally geometric uplift.

\section*{Acknowledgements}

CB is partially supported through the grants CEX2020-001007-S and PGC2018-095976-B-C21, funded by MCIN/AEI/10.13039/501100011033 and by ERDF A way of making Europe, and was supported by an FWO-Vlaanderen Postdoctoral Fellowship. 
SZ is supported by an FWO-Vlaanderen Doctoral Fellowship.
Both CB and SZ acknowledge the support of the FWO-Vlaanderen through the project G006119N and by the Vrije Universiteit Brussel through the Strategic Research Program ``High-Energy Physics''. 
We would like to thank Camille Eloy for helpful discussions.

\bibliography{CurrentBib}

\end{document}